\documentclass[12pt]{article}
\usepackage{graphicx}
\usepackage{latexsym}
\usepackage{amscd}
\usepackage[utf8]{inputenc}
\usepackage[T1, T2A]{fontenc}
\usepackage[english]{babel}
\usepackage{amsmath}
\usepackage{amssymb}
\usepackage{indentfirst}
\usepackage[small,centerlast]{caption2}
\usepackage{longtable}

\usepackage{hyperref}
\usepackage{color}
\usepackage{xcolor}

\let\vec=\mathbf

\makeatletter
\renewcommand{\section}{\@startsection {section}{1}{\z@}%
	{-3.5ex \@plus -1ex \@minus -.2ex}%
	{2.3ex \@plus.2ex}%
	{\normalfont\Large}}
\renewcommand{\subsection}{\@startsection{subsection}{2}{\z@}%
	{-3.25ex\@plus -1ex \@minus -.2ex}%
	{1.5ex \@plus .2ex}%
	{\normalfont\large\itshape}}
\renewcommand{\subsubsection}{\@startsection{subsubsection}{3}{1em}%
	{-3.25ex\@plus -1ex \@minus -.2ex}%
	{-1.5em \@plus .2em}%
	{\normalfont\normalsize\bfseries}}

\makeatother

\usepackage[%
backend=biber,% движок
bibencoding=utf8,% кодировка bib файла
sorting=none,% настройка сортировки списка литературы
style=gost-numeric,% стиль цитирования и библиографии (по ГОСТ)
autolang=other,% многоязычная библиография
sortcites=true,% сортировать номера затекстовых ссылок при цитировании 
movenames = false,% запрещает перемещение имен в область сведений об ответственности, если количество имен больше трех
maxnames = 3% отключает сокращение списка имен
]{biblatex}[2016/09/17]%
\DeclareFieldFormat[article]{title}{}
\DeclareFieldFormat[article]{titleaddon}{}

\DeclareFieldFormat[article]{journaltitle}{\mkbibemph{#1}}

\DefineBibliographyExtras{russian}{%

}
\DefineBibliographyExtras{english}{%

}   
\usepackage{xcolor}

\newcommand{\wpl}{\omega_\mathrm{p}}
\newcommand{\me}{m_\mathrm{e}}

\begin{document}

\begin{center}
	
	{\Large \textbf{Quasilinear interaction between Langmuir and Weibel turbulence in a beam–plasma system
    }}
	
	\medskip
	
	{\large \textbf{A.\:A.~Kuznetsov, Vl.\:V.~Kocharovsky}}
	
	\medskip
	\textit{A.V. Gaponov-Grekhov Institute of Applied Physics of the Russian Academy of Sciences, Nizhny Novgorod, Russia}\\

	\textit{kuznetsov.alexey@ipfran.ru}
	
\end{center}

\renewcommand{\abstractname}{}
\begin{abstract}
\noindent\small

To analyze the joint development of two-stream and filamentation kinetic instabilities in a plasma with a particle beam, a quasilinear approach has been developed that accounts for the integral nonlinear interaction of modes arising from the variation of the spatially averaged velocity distribution function of the particles. On this basis, a numerical study has been carried out within the initial two-dimensional problem for a range of characteristic parameters of the plasma and the beam, focusing on the evolution of Langmuir (two-stream) and Weibel (filamentation) turbulence spectra. It has been established that the evolving Weibel-type magnetic turbulence can significantly reshape the region of the velocity distribution that is resonant with Langmuir waves, thereby strongly influencing the formation and particularly the damping of Langmuir turbulence.
In turn, the Langmuir-type quasi-electrostatic turbulence can lead to substantial isotropization of the particle velocity distribution, thus altering the growth rates, evolution, and saturation levels of the Weibel turbulence modes.


\end{abstract}

\section{Introduction}

In various problems of laboratory and space collisionless plasma physics, including the solar wind and the magnetospheres of stars and planets, the presence of a beam of energetic charged particles (electrons and/or ions) propagating through a warm plasma is a typical feature~\cite{Gary1993,Treumann1997,Marsch2006}. Even if the velocity distributions of the particles in both the plasma and the beam are isotropic Maxwellian (in their respective rest frames), the beam–plasma system is characterized by an anisotropic velocity distribution. As a result, according to dispersion relation analysis~\cite{Mikhailovsky1971,Fried1959,Krall1973,Tzoufras2006,Bret2010}, various instabilities can develop simultaneously—most notably, the resonant quasi-electrostatic two-stream instability and the aperiodic quasi-magnetostatic Weibel-type instability, respectively termed the beam and filamentation instabilities.

The two-stream instability manifests through the growth of longitudinal Langmuir modes, leading to the formation of short-wavelength Langmuir turbulence in the electric field and plasma density~\cite{Vedenov1963,Zakharov1972,VedenovRyutov1975,Krall1973,Appert1976,Yi2010,Bakunin2017,Sun2022}. As a result, a plateau emerges in the particle velocity distribution in the region between the thermal plasma and the beam.
The filamentation instability, in turn, gives rise to Weibel-type (magnetic) turbulence, characterized by the generation of longer-wavelength transverse quasi-magnetostatic fields accompanied by corresponding current filaments or layers~\cite{Weibel1959, Zhou2022,Fried1959, Kalman1968, Morse1971, Kocharovsky2016, Lazar2006, Stockem2009, SchaeferRolffs2006}. This process reduces the anisotropy of the total particle velocity distribution and smooths the beam-related region of the distribution.

We focus on an unmagnetized plasma, for which the two-stream (beam-type) and filamentation (Weibel-type) instabilities have been extensively studied individually, particularly within the linear approximation (see, e.g.,\cite{Mikhailovsky1971,Fried1959,Tzoufras2006,Hao2008,Bret2010,Moya2022}). At the same time, to investigate the long-term nonlinear dynamics of the turbulence arising from the development of isolated two-stream or filamentation instabilities, both fully kinetic particle-in-cell (PIC) simulations~\cite{Kasaba2001,Dum1994,Yi2010,Ruyer2015,Dieckmann2009,Bret2010,Lazar2023,Nechaev2023,Kocharovsky2024,Garasev2022,Korzhimanov2024,Kuznetsov2025a} and various approximate approaches—primarily quasilinear theory—have been employed~\cite{Vedenov1963,VedenovRyutov1975,Appert1976,Bakunin2017,Ziebell2008,Lemons1979,Davidson1972,Ruyer2015,Kuznetsov2022,Kuznetsov2023}.

However, the problem of nonlinear interaction between Langmuir and Weibel turbulence remains poorly explored, despite its importance for understanding a wide range of phenomena, such as the formation of collisionless shock waves and structures in accretion disks, the interpenetration of adjacent clouds and particle streams in the stellar wind, the evolution of coronal mass ejections and solar flares, and the heating of plasma and modification of its kinetic properties during the injection of high-energy particle beams, among others~\cite{Marcowith2016,Treumann2015,Aschwanden2005,Medvedev2006,Nishikawa2009,Kato2007,Kuznetsov2025b}. Existing studies in this area are mostly based on particle-in-cell (PIC) simulations, often constrained to the hydrodynamic regime of a particular instability, and generally do not address, much less provide a detailed analysis of, the mutual influence of Langmuir and Weibel turbulence; see, e.g.,\cite{Kong2009,Ruyer2015,Bret2010,Lazar2023} (for the magnetized case, see also~\cite{Lazar2023,Lopez2020}). For example, the review~\cite{Bret2010}, which is arguably the most comprehensive source relevant to the unmagnetized case considered here, discusses saturation mechanisms and possible nonlinear localized structures associated with each type of turbulence individually, but does not investigate how turbulence of one type modifies the particle distribution function and thereby influences the development of the other~--- although the possibility of their sequential development is mentioned.

In the present work, we undertake a detailed study of one of the mechanisms underlying the mutual influence between the two aforementioned instabilities, focusing on the kinetic regime of their growth, saturation, and the subsequent relaxation of the resulting weak turbulence. Specifically, we analyze how each of the turbulent fields~--— primarily electric in the case of the two-stream instability and magnetic in the case of the filamentation instability~--— modifies the spatially averaged velocity distribution function of particles, thereby altering the evolution of the accompanying instability. Within such a quasilinear framework~\cite{VedenovRyutov1975,Appert1976,Kuznetsov2022,Kuznetsov2023}, each turbulence mode evolves locally in time according to its instantaneous linear growth (or damping) rate, which is calculated using the current velocity distribution function. The analysis is restricted to regimes where, as confirmed by test particle-in-cell simulations, other possible mechanisms of nonlinear spectral evolution—such as the formation of multiple Langmuir solitons with uncompensated charge, current z-pinches, or systematic three- and four-wave interactions among individual modes (spatial harmonics)~--- can be ruled out as dominant.

To this end, a closed quasilinear system of equations has been formulated, and a novel numerical code has been developed to solve it. Using this code, we have computed the simultaneous nonlinear evolution of a large number (several thousands) of Langmuir and Weibel modes in a turbulent plasma under the kinetic approximation, in which the growth rates $\gamma$ and corresponding wavenumbers $k$ of all modes satisfy the condition $\gamma \ll k v_T$, where $v_T$ is the thermal velocity. As a result, the role of quasilinear interactions in the evolution of Langmuir and Weibel turbulence spectra has been established. In addition, we have obtained the temporal dynamics of the root-mean-square electric and magnetic fields and demonstrated the nonuniform character of both the reduction of the velocity-space anisotropy of the particle distribution function and the increase of plasma thermal energy (heating) in the direction transverse to the beam.

This approach is consistent with the classical quasilinear description of Langmuir turbulence~\cite{VedenovRyutov1975,Appert1976}, as well as with the formulation developed in~\cite{Kuznetsov2022,Kuznetsov2023} for Weibel-type turbulence. Compared to particle-in-cell (PIC) simulations, it offers several advantages, including the ability to freely select the range of wave vectors of interest and to control the initial amplitude levels of the modes.

This flexibility enables not only the independent verification of the quasilinear treatment of filamentation and two-stream modes by comparing specific outcomes to previously reported PIC results, but also the analysis of their individual spectral evolution in contrast to their joint evolution. This allows not only for independent verification of the quasilinear treatment of filamentation and two-stream instabilities~--- by comparing spectral characteristics to previously reported PIC results~--- but also for a comparative analysis of their separate spectral evolution versus their joint nonlinear dynamics. An additional advantage is the significantly lower computational cost of the quasilinear method, particularly in contrast to PIC simulations, which require considerably more resources due to the need for a large number of particles to accurately describe resonance effects associated with the kinetic two-stream instability~\cite{Lotov2015}.

\begin{figure}[h]
\setcaptionmargin{5mm}
\includegraphics[width=0.5\linewidth]{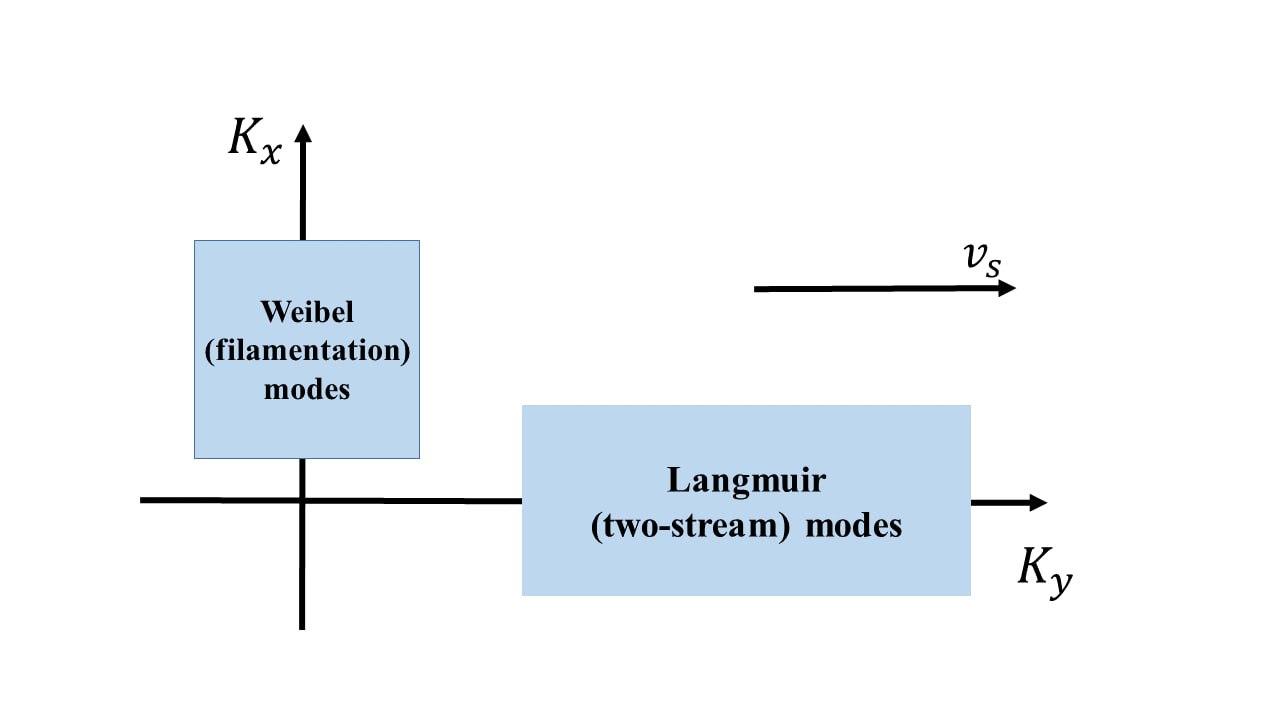}
\centering
\captionstyle{normal}
\caption{Schematic of the computational domain, showing two rectangular grids in wavevector space corresponding to the generated spectra of filamentation and two-stream modes.}
\label{fig:1}
\end{figure}

For simplicity, the results presented here are limited to the initial-value two-dimensional problem for a homogeneous, unmagnetized, nonrelativistic plasma–beam system. In this configuration, both the background plasma and the beam consist of cold ions and electrons with the same thermal velocity $v_T$ (and temperature $T$), while the beam propagates along the $y$-axis with a drift velocity $v_s$ that exceeds the thermal velocity by several times. Under these assumptions, the development of localized space charge and current-induced $z$-pinches is excluded~\cite{Tzoufras2006,Hao2008,Kocharovsky2024,Garasev2022}. The study focuses on the quasilinear evolution of the spectrum of two-stream (Langmuir) modes~--- essentially electrostatic perturbations with wave vectors predominantly aligned with the beam~--- and that of filamentation (Weibel) modes~--- predominantly transverse electromagnetic perturbations with wave vectors nearly orthogonal to the beam direction~\cite{Bret2004,Bret2010}. These two regions of unstable modes in wave-vector space are well separated in both orientation and wavenumber range, and are represented using two distinct rectangular grids that densely cover the corresponding instability domains (Fig.~\ref{fig:1}).

The quasilinear system of equations employed for the simulations is presented in Section 2. Sections 3 and 4 discuss the results of numerical calculations for two representative sets of plasma and beam parameters, which exhibit qualitatively distinct dynamics of the spectrum
of modes. The conclusions summarize the key findings and outline potential directions for future research.

\section{Quasilinear system of equations}

For a collisionless plasma, in which the motion of heavy ions can be neglected over the characteristic evolution timescale of Weibel turbulence~--— on the order of ten instability saturation times~--— the self-consistent Vlasov–Maxwell equations for the electron distribution function $f(v_x, v_y, x, y, t)$, which includes the electron density $N(x, y, t)$, and for the electric field $\vec{E} = (E_x, E_y, 0)$ and magnetic field $\vec{B} = (0, 0, B_z)$, take the form~\cite{Mikhailovsky1971,Krall1973,Baumjohann2012,Vedenov1963,VedenovRyutov1975,Bakunin2017,Kocharovsky2016}:
\begin{eqnarray}
    \dfrac{\partial f}{\partial t}+\vec{v}\dfrac{\partial f}{\partial \vec{r}}+\dfrac{e}{\me} \left(\vec{E}+\dfrac{1}{c}\left[\vec{v},\vec{B}\right]\right) \dfrac{\partial f}{\partial \vec{v}}=0, \\
    \label{eq:maxw1} 
    \nabla \times \vec{B}=\dfrac{1}{c}\dfrac{\partial \vec{E}}{\partial t}+\dfrac{4\pi}{c}\vec{j}, \\
    \label{eq:maxw2}
    \nabla \times \vec{E}=-\dfrac{1}{c}\dfrac{\partial \vec{B}}{\partial t},
\end{eqnarray}
where $c$ is the speed of light in vacuum, $e$ and $m_e$ are the electron charge and mass, respectively; $\vec{j} = e \iint_{-\infty}^{+\infty} \vec{v} f(v_x, v_y, x, y, t) , dv_x dv_y$ is the current density; and $N = \iint_{-\infty}^{+\infty} f(v_x, v_y, x, y, t) , dv_x dv_y$ is the electron number density. It is assumed that both the spatial and velocity vectors have only two nonzero components, $\vec{r} = (x, y, 0)$ and $\vec{v} = (v_x, v_y, 0)$, in accordance with the two-dimensional formulation of the problem.

When a sufficiently wide particle beam, with a transverse size much greater than the plasma skin depth, propagates through a background plasma, a return current is induced such that $n_b \beta_b = n_s \beta_s$~\cite{Shukla2018,Jia2013,Karlick2008}, where $n_i$ and $\beta_i = v_i/c$ denote the initial density and dimensionless directed velocity of each population; the indices $b$ and $s$ correspond to the background plasma and the beam, respectively. The dimensionless initial electron velocity distribution used in this work is thus given by
\begin{equation}
\label{eq:bp}
    \Psi_0(\vec{\beta})=\dfrac{n_b}{\pi\beta_T^2(n_b+n_s)} \exp\left(-\dfrac{\beta_x^2}{\beta_T^2}-\dfrac{\left(\beta_y+\beta_{b}\right)^2}{\beta_T^2}\right)+\dfrac{n_s}{\pi\beta_T^2 (n_b+n_s)} \exp\left(-\dfrac{\beta_x^2}{\beta_T^2}-\dfrac{\left(\beta_y-\beta_{s}\right)^2}{\beta_T^2}\right).  
\end{equation}
where $\beta_{x,y} = v_{x,y}/c$, i.e., $\vec{\beta} = \vec{v}/c$, and $\beta_T = \sqrt{2 n k_b T / m_e}/c$, with $k_b$ denoting Boltzmann’s constant.

The quasilinear approach to modeling the interplay between Langmuir and Weibel turbulence is based on a spectral expansion of the Vlasov–Maxwell equations~\cite{Baumjohann2012}, initialized with a broadband, noise-like perturbation of the electromagnetic field across the spectral domains associated with filamentation and two-stream instabilities. This expansion is particularly effective when the dominant nonlinear interaction mechanism consists in the collective, integral variation of the velocity distribution function by the entire spectrum of modes. The evolving distribution function determines the instantaneous values of the growth (or damping) rates and real frequencies for all individual modes, which otherwise evolve independently. Such a quasilinear description is justified under the random phase approximation for the modes and in the absence of particle trapping by the electric or magnetic fields of individual harmonics or localized structures. These assumptions typically hold in the kinetic regime of instability development and during the subsequent weakly turbulent nonlinear evolution~\cite{GaleevSagdeev1969,Bakunin2017,Kuznetsov2023}. The presence of a large number of incoherent modes that are densely spaced and span the entire significant region of instability ensures smoothness in the velocity-space distribution function and prevents the emergence of artificial oscillations or negative values due to coherent interference, except possibly within negligible regions of velocity space far beyond the thermal range. In this regime, the dynamics of each individual mode is essentially adiabatic, continuously adjusting to the evolving electron distribution function.

In the two-dimensional inhomogeneous case considered here, Weibel turbulence is modeled using a grid of wave vectors consisting of $m_W \cdot s_W$ non-collinear modes with chaotic phases and approximately equal initial amplitudes at $t=0$, that are densely spaced and span the entire significant region of instability. A similar grid of $m_L \cdot s_L$ non-collinear modes is employed to describe Langmuir turbulence as well (see Fig.~\ref{fig:1}), where $m_i$ and $s_i$ denote the numbers of discrete values assigned to the corresponding orthogonal projections $\vec{K}\vec{x_0}$ and $\vec{K}\vec{y_0}$ of the wave vectors. In this representation, the magnetic field is expressed as a sum over the integer-valued vector index $\vec{n} = (n_x, n_y)$:

\begin{equation}
    B_z(t,x,y)= \mathrm{Re} \Biggr[ \sum^{m_W,s_W}_{n_x,n_y=1}B_{k_\vec{n}}(t)\exp(- ik_{n_{x}}x - ik_{n_{y}}y)\Biggr]+ \mathrm{Re} \Biggr[\sum^{m_L,s_L}_{n_x,n_y=1}B_{k_\vec{n}}(t)\exp(- ik_{n_{x}}x - ik_{n_{y}}y)\Biggr].
\end{equation}

Analogous expansions apply to both components of the electric field, $E_x$ and $E_y$, as well as to the perturbation of the distribution function $\delta f(t, x, y)$. The resulting quasilinear system derived from the Vlasov–Maxwell equations consists of $4(m_W \cdot s_W + m_L \cdot s_L) + 1$ equations for the spatially averaged electron velocity distribution function $\psi_0$, the dimensionless complex amplitudes of its perturbations $\psi_{K_{\vec{n}}}$, and the perturbations of the magnetic field $b_{K_{\vec{n}}}$ and the two components of the electric field $e_{x{K_{\vec{n}}}}$ and $e_{y{K_{\vec{n}}}}$, and is given by the following set of equations:

\begin{equation}
\label{eq14}
    \dfrac{\partial \psi_0}{\partial \tau} 
    + \mathrm{Re}\Biggr[\sum\limits^{m_W,s_W}_{n_x,n_y=1} \hat \Phi(b_{K_{\vec{n}}},\overrightarrow{e}_{K_{\vec{n}}},\psi_{K_{\vec{n}}}^*) 
     \Biggr]+ \mathrm{Re}\Biggr[\sum\limits^{m_L,s_L}_{n_x,n_y=1} \hat \Phi(b_{K_{\vec{n}}},\overrightarrow{e}_{K_{\vec{n}}},\psi_{K_{\vec{n}}}^*) 
     \Biggr]=0,
\end{equation}
\begin{equation}
    \label{eq15}
    \dfrac{\partial \psi_{K_{\vec{n}}}}{\partial \tau}+iK_{n_x}\beta_x\psi_{K_{\vec{n}}}+iK_{n_y}\beta_y\psi_{K_{\vec{n}}}+2\hat \Phi(b_{K_{\vec{n}}},\psi_0)=0,
\end{equation}
\begin{equation}
    \dfrac{\partial b_{K_{\vec{n}}}}{\partial \tau}=-ie_{y{K_{\vec{n}}}}K_{n_x}+ie_{x{K_{\vec{n}}}}K_{n_y},
\end{equation}
\begin{equation}
    \dfrac{\partial e_{x{K_{\vec{n}}}}}{\partial \tau}=ib_{K_{\vec{n}}}K_{n_y}-\beta_{\|}^{-1}\iint\limits^{+\infty}_{-\infty}\beta_x\psi_{K_{\vec{n}}}(\tau,\beta_x,\beta_y)d\beta_xd\beta_y,
\end{equation}
\begin{equation}
\label{eq19}
    \dfrac{\partial e_{y{K_{\vec{n}}}}}{\partial \tau}=-ib_{K_{\vec{n}}}K_{n_x}+\beta_{\|}^{-1}{\iint\limits^{+\infty}_{-\infty}\beta_y\psi_{K_{\vec{n}}}(\tau,\beta_x,\beta_y)d\beta_xd\beta_y} .
\end{equation}

Dimensionless time and wave number are used here, as well as normalized (complex) modes of the magnetic and electric field and the electron velocity distribution function. 
\begin{equation}
    \label{eq19plus1}
    \tau=\wpl t, \
    K=\dfrac{kc}{\wpl}; \ 
    \wpl^2=\dfrac{4\pi Ne^2}{\me},\
    b_{K_{\vec{n}}}=\dfrac{B_{K_{\vec{n}}}}{\sqrt{8\pi N T}},\
    T=\dfrac{m_ec^2\beta_{T}^2}{2};\
    \psi_{K_{\vec{n}}}=\dfrac{c^2f_{ K_{\vec{n}}}}{N}.\ 
\end{equation}

The complex components of the electric field $e_{x{K_{\vec{n}}}}$ and $e_{y{K_{\vec{n}}}}$ are normalized in the same way as the magnetic field $b_{K_{\vec{n}}}$. To simplify the expressions in Eqs.~(\ref{eq14})–(\ref{eq15}), a differential operator is introduced:

\begin{equation}
\label{eq:operator}
    \hat \Phi(b_{K_n},e_{x{K_{\vec{n}}}},e_{y{K_{\vec{n}}}},\psi(\vec{\beta}))=\dfrac{e_{y{K_{\vec{n}}}}}{2}\dfrac{\partial \psi(\vec{\beta})}{\partial \beta_y}+\dfrac{e_{x{K_{\vec{n}}}}}{2}\dfrac{\partial \psi(\vec{\beta})}{\partial \beta_x}-\dfrac{b_{K_n}}{2} \left(\beta_x\dfrac{\partial \psi(\vec{\beta})}{\partial \beta_y}-\beta_y\dfrac{\partial \psi(\vec{\beta})}{\partial \beta_x}\right).
\end{equation}

The  system of equations~(\ref{eq14})–(\ref{eq19}), together with the operator defined in~(\ref{eq:operator}), was solved numerically using the standard Störmer–Verlet (Leapfrog) method~\cite{Birdsall2018}. The anisotropy parameter $A$, defined as the deviation from unity of the ratio of the total electron energy in the longitudinal direction $W_\parallel$ (along the beam velocity) to that in the transverse direction $W_\perp$, plays a key role in the quasilinear description of Weibel turbulence~\cite{Weibel1959,Davidson1972,Lemons1979,Tzoufras2006,Kuznetsov2023}. The initial value of the anisotropy parameter $A_0$ for the electron distribution given by Eq.~(\ref{eq:bp}) is determined by the initial ratio between the directed and thermal energy components:

\begin{equation}
A=\frac{W_\|}{W_\perp}-1;~~~A(\tau=0)=A_0=\frac{n_s\beta_s^2+n_b\beta_b^2}{(n_s+n_b)\beta_T^2/2}.
\label{anisotropy}
\end{equation}

The time step $d\tau$ and the velocity-space grid step $d\beta$ used to approximate the electron distribution were chosen to be at least an order of magnitude smaller than the smallest characteristic timescale and velocity scale, respectively. The total number of spatial modes, $m_W \cdot s_W + m_L \cdot s_L$, associated with both filamentation and Langmuir turbulence, typically amounted to several thousand in the simulations and was selected to satisfy two conditions: (i) that the root-mean-square values of the magnetic and electric fields remained insensitive (within a few percent) to further increases in the number of modes, and (ii) that resonant overlap was achieved~\cite{GaleevSagdeev1969,Bakunin2017} for energy-carrying Langmuir modes throughout the entire interval of nonlinear evolution under study. In all simulations, the validity of the kinetic approximation was verified by ensuring that the growth rate satisfied $\gamma \ll K \beta_T$, where the initial electron thermal velocity was fixed to $\beta_T = 0.1$ for definiteness.

\section{Quasilinear suppression of Langmuir turbulence by filamentation modes}

Although the two-stream instability is primarily driven by a relatively small population of resonant electrons~---  since the exponential growth of quasi-electrostatic fields results from the presence of a localized nonmonotonic region in the electron velocity distribution~--- its growth rate for nonrelativistic beams with directed velocity $\beta_s$ significantly exceeding the thermal velocity $\beta_{T}$ is typically much greater than that of the filamentation instability~\cite{Bret2010}. The amplitude of Langmuir waves increases during the development of the two-stream instability until the aforementioned nonmonotonicity is smoothed out and a quasilinear plateau is formed. As a result, part of the energy associated with the beam-directed motion of electrons is redistributed between the generated fields and the thermal energy of the electrons, leading to a partial isotropization of their velocity distribution~\cite{GaleevSagdeev1969}. In the one-dimensional approximation, the quasilinear dynamics of Langmuir turbulence reaches a quasi-stationary state at this stage. However, in the considered two-dimensional case, the plateau continues to evolve due to Landau damping, primarily affecting oblique modes that are not collinear with the beam, so that the spectrum of quasi-electrostatic turbulence gradually shifts toward shorter wavelengths, and the beam-induced anisotropy in the velocity distribution is progressively diminished~\cite{Appert1976,Yi2010}. Thus, in the absence of magnetic turbulence, the beam–plasma system undergoes a very slow, quasilinear isotropization of the particle distribution and a gradual decay of the electric field.

In the absence of space-charge effects~--- ensured by the assumption of equal background and beam temperatures~--- any anisotropic electron distribution of the form~(\ref{eq:bp}) is subject to the filamentation instability~\cite{Tzoufras2006}. Therefore, under typical collisionless conditions, the filamentation instability in the  beam–plasma system is inevitable and continues to develop even after the saturation of the two-stream instability. However, the reduction of the electron distribution anisotropy parameter due to Langmuir turbulence suppresses the growth rate of filamentation modes and lowers the saturation level of their evolution.  In turn, the resulting magnetic turbulence driven by the filamentation modes can, as will be shown later, lead to a significant modification of the electron velocity distribution in the region resonant with Langmuir waves and, consequently, strongly influence their Landau damping.

\begin{figure}[h]
\setcaptionmargin{5mm}
\includegraphics[width=0.5\linewidth]{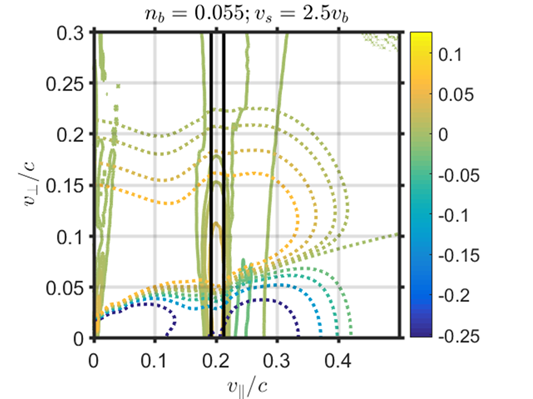}
\centering
\captionstyle{normal}
\caption{Contours (at $\omega_p t = 5000$) of the correction to the spatially averaged electron velocity distribution function, normalized to the peak of the initial distribution (\ref{eq:bp}), which at $t=0$ was a Maxwellian with a beam. Solid lines show the case with two-stream modes only, while dashed lines correspond to the joint evolution of Langmuir and Weibel turbulence. Initial distribution parameters (Eq. 4): $n_s=0.055n_b$, $\beta_s=2.5\beta_T$, $\beta_T=0.1$. Vertical black lines indicate the estimated resonance region for beam-aligned Langmuir modes, which contain the bulk of the energy at this time.
}
\label{fig:FR1}
\end{figure}

To illustrate the above, let us consider the results of a quasilinear simulation of the system evolution with initial parameters $n_s = 0.055n_b$ and $\beta_s = 2.5\beta_T$. According to linear theory~\cite{Mikhailovsky1971,Davidson1972}, the fastest-growing longitudinal Langmuir mode has a growth rate of $\gamma_L \approx 0.017\omega_p$ and a wave number $K_y \approx 4.2$, while the most unstable filamentation mode corresponds to a transverse wave number $K_x \approx 0.44$, with a growth rate of $\gamma_F \approx 9 \times 10^{-3}\omega_p$, approximately half that of the Langmuir mode. For this choice of initial particle distribution, a comparison between the distribution function corrections obtained in simulations that include only two-stream modes and those accounting for the coupled dynamics of Langmuir and Weibel turbulence reveals that the impact of filamentation modes on the electron distribution is significantly greater than that of the Langmuir modes. Moreover, the modification of the distribution function caused by filamentation turbulence extends into the resonance region associated with beam-aligned Langmuir modes, which are the dominant energy-carrying modes at that stage of evolution (see Fig.~\ref{fig:FR1}). The resonance interval boundaries shown in the figure, $\beta_{y,\min} = 1/K_{y,\max}$ and $\beta_{y,\max} = 1/K_{y,\min}$, are estimated from the Cherenkov resonance condition, under the simplifying assumption that the Langmuir wave frequency equals the plasma frequency and the transverse wave number component is zero, i.e., $K_x = 0$.

\begin{figure}[h]
\setcaptionmargin{5mm}
\includegraphics[width=0.7\linewidth]{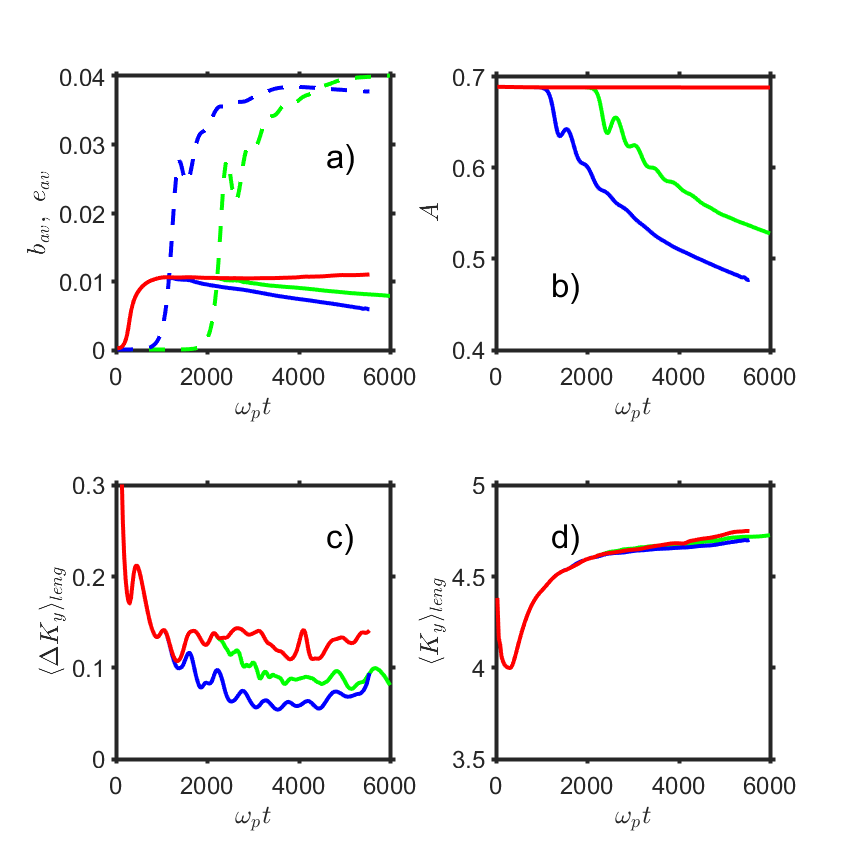}
\centering
\captionstyle{normal}
\caption{Temporal evolution of (a) the root-mean-square magnetic field $b_{av}$ of filamentation modes (dashed lines) and the electric field $e_{av}$ of two-stream modes (solid lines), (b) the anisotropy parameter $A$, (c) the characteristic spectral width $\langle \Delta K_y \rangle$ of Langmuir waves along the beam direction, and (d) their characteristic longitudinal wavenumber $\langle K_y \rangle$, for three cases: two-stream modes only (red), and joint evolution with filamentation modes having a $10^4$ difference in their initial amplitudes (blue and green), while the initial amplitudes of the two-stream modes are identical. Initial distribution parameters (\ref{eq:bp}): $n_s=0.055n_b$, $\beta_s=2.5\beta_T$, $\beta_T=0.1$.}
\label{fig:average1}
\end{figure}

In the absence of filamentation modes, the root-mean-square electric field of Langmuir modes remains approximately constant over time intervals not exceeding six times the saturation time of the two-stream instability. When filamentation modes are included, their exponential growth—culminating in the strongest harmonics—results in a significant anisotropic modification of the overall electron velocity distribution function. This transition to the nonlinear stage of magnetic turbulence leads to a rapid decay of Langmuir waves (see Fig.\ref{fig:average1}a). The damping rate of each Langmuir mode is governed exclusively by the shape of the velocity distribution function within its resonance region. As the Langmuir wave spectrum gradually shifts toward shorter wavelengths over time (see Fig.\ref{fig:average1}d), the decay rate of the root-mean-square Langmuir electric field becomes sensitive to the time interval between the saturation of the two-stream and filamentation instabilities, and thus to the initial amplitudes of the electric and magnetic field perturbations (illustrated in Fig.~\ref{fig:average1}a). Throughout the entire evolution, the root-mean-square inductive electric field associated with filamentation modes remains small compared to that of Langmuir modes. Likewise, the root-mean-square magnetic field of Langmuir modes is negligible in comparison to the field generated by filamentation modes.

In the absence of filamentation modes, the root-mean-square electric field associated with Langmuir modes remains approximately constant over time intervals not exceeding six times the saturation time of the two-stream instability. The inclusion of filamentation modes results in a pronounced damping of Langmuir waves that begins immediately after the end of their exponential growth, which marks the onset of substantial anisotropic deformation of the total electron velocity distribution function and the transition to the nonlinear stage of magnetic turbulence (see Fig.~\ref{fig:average1}a). The damping rate of each individual Langmuir mode is determined solely by the shape of the velocity distribution function in its corresponding resonance region. As the Langmuir wave spectrum progressively shifts toward shorter wavelengths over time (see Fig.\ref{fig:average1}d), the decay rate of the root-mean-square Langmuir electric field becomes dependent on the time interval between the saturation of the two-stream and filamentation instabilities, and therefore on the initial amplitudes of the electric and magnetic field perturbations (as illustrated in Fig.~\ref{fig:average1}a). Throughout the entire evolution, the root-mean-square inductive electric field of filamentation modes remains small compared to that of Langmuir modes. Similarly, the root-mean-square magnetic field of Langmuir modes is negligible in comparison to the magnetic field generated by filamentation modes.

As noted above, due to quasilinear interactions among Langmuir waves, the characteristic parallel wavenumber of their spectrum, $\langle K_y\rangle$, increases over time (see Fig.\ref{fig:average1}d). In the case considered here, this increase saturates at a level of about 20\% roughly one order of magnitude after the saturation time of Langmuir turbulence and is almost unaffected by the presence of Weibel turbulence. However, the latter does influence the quasilinear deformation of the plateau region in the resonant part of the beam electron velocity distribution, which leads to a significant narrowing of the longitudinal characteristic spectral width of Langmuir waves, $\langle \Delta K_y\rangle$, by a factor of approximately 1.5 (see Fig.\ref{fig:average1}c), as a result of the relatively rapid damping of the shortest-wavelength part of the spectrum. In contrast, the transverse characteristic spectral width, $\langle \Delta K_x\rangle$, changes only slightly, within a few percent.

The dynamics of Weibel turbulence, which develops at substantially longer wavelengths, is governed by non-resonant wave–particle interactions~\cite{GaleevSagdeev1969,Kuznetsov2023} and is weakly sensitive to the form of the electron velocity distribution in the resonant region associated with Langmuir waves. A key parameter in the quasilinear description of this turbulence is the integral anisotropy parameter $A$, defined as the difference from unity of the ratio of the total electron energy in the parallel ($W_|$) and perpendicular ($W_\perp$) directions with respect to the direction of the beam velocity~(\ref{anisotropy}). As the turbulence evolves, the electron distribution becomes increasingly isotropic, which is indicated by a monotonic decrease of the anisotropy parameter (see Fig.~\ref{fig:average1}b).

In the case under consideration, Langmuir turbulence proves to be negligible for the growth of magnetic field fluctuations, and the reduction of the anisotropy parameter is governed solely by Weibel turbulence. At its saturation stage, this reduction occurs quite rapidly and reaches approximately 30\% of the initial value $A_0=0.687$. Subsequently, the anisotropy parameter decreases significantly more slowly, while the spectrum of Weibel turbulence shifts self-similarly toward the long-wavelength region. At the same time, the root-mean-square magnetic field $b_{av}$, immediately after the end of the exponential growth phase, exhibits an approximately power-law increase during a short intermediate stage, reaches a maximum, and then decays slowly quasi-linearly (Fig.\ref{fig:average1}a)~\cite{Kuznetsov2023}. Throughout this entire entire stage, the magnetic energy exceeds the energy of the quasi-electrostatic field by at least an order of magnitude.

This spectral behavior is fully consistent with previously obtained results from the quasilinear approach to the evolution of an initially bi-Maxwellian electron distribution in the limit of low initial anisotropy, where the mean wave number of the spectrum $\langle K_x\rangle$ decreases approximately according to a power-law dependence~\cite{Kuznetsov2023,Borodachev2016_Radiofiz}. A similar pattern was observed in all quasilinear simulations performed for the beam-plasma system with a small initial anisotropy parameter ($A_0\lesssim1$) and a high level of resulting magnetic turbulence compared to the Langmuir turbulence.

The heating of the plasma, which in this context refers to the increase in the transverse thermal energy of electrons, $W_\perp$, is a non-uniform process that occurs in three distinct stages:

1) dissipation of quasi-electrostatic fields via Landau damping during the time interval between the saturation of Langmuir turbulence and the onset of Weibel turbulence saturation;

2) predominant dissipation of quasi-magnetostatic fields, which is especially efficient during their saturation stage, when the mode amplitudes are large and vary rapidly, generating the strongest inductive electric fields;

3) slower dissipation of both types of quasi-static fields at times significantly exceeding the saturation time of magnetic turbulence.

The third stage is not shown in Fig.~\ref{fig:average1}, and the first stage is only weakly expressed due to the low level of developed Langmuir turbulence. These heating stages also occur in the case presented in the next section, where the level of Langmuir turbulence is higher and, consequently, the heating during the first stage is much more pronounced.

\section{Attenuation of magnetic turbulence development by Langmuir modes}

At lower beam velocities than those considered in the previous section, the growth rate of two-stream modes becomes weaker, and these modes may develop simultaneously with filamentation modes. However, the latter still dominate in amplitude and suppress the former, remaining effectively unaffected by the presence of very weak Langmuir turbulence. Therefore, we turn to the analysis of a case with a higher beam velocity compared to the previous section, in which the conditions are such that the influence of Langmuir turbulence on filamentation modes becomes significant, and the contributions of both types of turbulence to the isotropization of the electron velocity distribution~--- that is, to the reduction of the anisotropy parameter $A$~--- become comparable.

\begin{figure}[h]
\setcaptionmargin{5mm}
\includegraphics[width=0.7\linewidth]{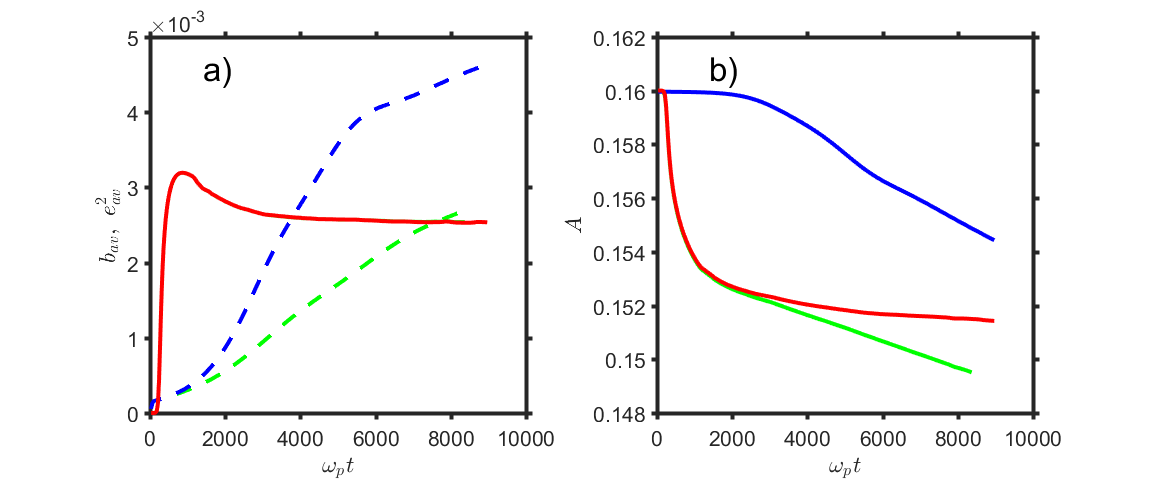}
\centering
\captionstyle{normal}
\caption{
Temporal evolution of (a) the root-mean-square magnetic field $b_{av}$ of filamentation modes (dashed lines) and the mean square electric field $e_{av}^2$ of two-stream modes (solid lines), and (b) the anisotropy parameter $A$, for three scenarios: two-stream modes only (red), filamentation modes only (blue), and their joint evolution (green). Initial distribution parameters (\ref{eq:bp}): $n_s=0.005n_b$, $\beta_s=4\beta_T$, $\beta_T=0.1$.} 
\label{fig:average_v4}
\end{figure}

\begin{figure}[h]
\setcaptionmargin{5mm}
\includegraphics[width=1\linewidth]{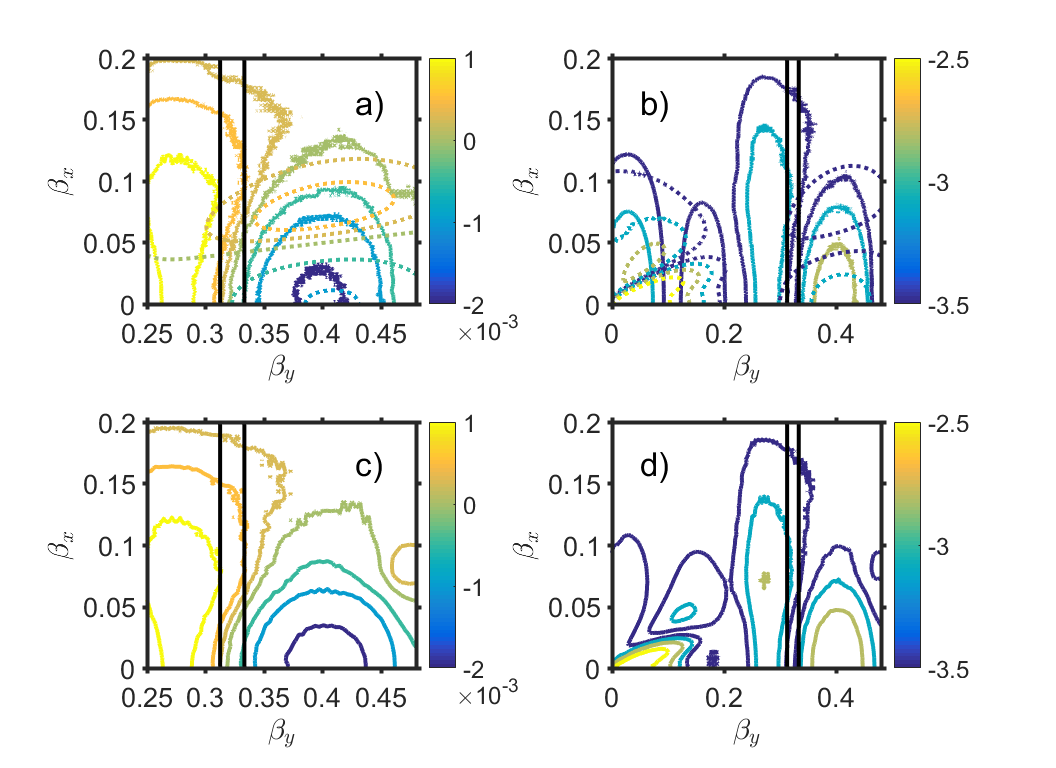}
\centering
\captionstyle{normal}
\caption{(a) Contours (at $\omega_p t = 8100$) of the correction to the spatially averaged electron velocity distribution function, normalized to the peak of the initial Maxwellian-with-beam distribution (\ref{eq:bp}), for the case with two-stream modes only (solid) and with filamentation modes only (dashed). Initial distribution parameters (\ref{eq:bp}): $n_s=0.005n_b$, $\beta_s=4\beta_T$, $\beta_T=0.1$. Vertical black lines indicate the estimated resonance region for beam-aligned Langmuir modes, which contain the bulk of the energy at this time.
(b) Same as (a), but for the base-10 logarithm of the correction to the distribution function.
(c)–(d) Same as (a)–(b), respectively, for the case with joint evolution of two-stream and filamentation modes.
}
\label{fig:fr2}
\end{figure}

Specifically, we consider a beam with approximately twice the directed velocity and an order of magnitude lower particle density, $n_s = 0.005n_b$, $\beta_s = 4\beta_T$, while maintaining a small initial anisotropy, $A_0 = 0.16$, and a moderate two-stream growth rate, only about an order of magnitude higher than that of the filamentation modes. As shown in Fig.\ref{fig:average_v4}, similar to the previous case, the two-stream instability reaches saturation earlier, since the increment of the most unstable longitudinal mode with wave number $K_y \approx 3.2$ is $\gamma_L \approx 0.027\omega_p$, whereas the most unstable transverse filamentation mode with wave number $K_x \approx 0.226$ has an increment of only $\gamma_W \approx 1.27 \times 10^{-3}\omega_p$. As a result, the emerging Langmuir turbulence alters the electron velocity distribution in the region between the background plasma and the beam (see Fig.\ref{fig:fr2}), leading to a noticeable reduction of the quasilinear growth rates of filamentation modes and to an approximately twofold decrease in the saturation level of the root-mean-square magnetic field of the Weibel turbulence (Fig.~\ref{fig:average_v4}a). At the same time, spectral characteristics such as the mean wave number $\langle K\rangle$, as well as the transverse $\langle\Delta K_x\rangle$ and longitudinal $\langle \Delta K_y\rangle$ widths of the Weibel spectrum, are almost not affected by the presence of Langmuir turbulence, differing by no more than 1–2\% between simulations with and without two-stream modes included.

Over the considered timescale, the additional isotropization of the electron distribution by filamentation modes occurs predominantly outside the velocity region resonant with Langmuir waves~(see Fig.\ref{fig:fr2}) and, in fact, begins only upon the transition to the saturation stage of the Weibel turbulence at $\omega_p t > 5000$ (cf. the red and green curves for the anisotropy parameter in Fig.\ref{fig:average_v4}b). As a result, in simulations with and without the inclusion of filamentation modes, the dynamics of the Langmuir turbulence~--- specifically, the root-mean-square electric field, characteristic wave number, and both transverse and longitudinal spectral widths~--- differ by no more than 10\%. This, however, does not eliminate the necessity of accounting for filamentation modes in the analysis of Langmuir wave damping at later times, at which point the quasilinear description may become invalid due to the breakdown of its applicability conditions, for example, as a result of the accumulation of nonlinear three- or four-wave interaction effects between individual modes.

\section{Conclusion}

The results presented above illustrate possible scenarios of quasilinear interaction between evolving spectra of Langmuir and Weibel turbulence in a collisionless, unmagnetized, nonrelativistic plasma consisting of a background population and a warm electron beam. The initial electron distribution is assumed to be Maxwellian with the same temperature for both components, while ions are considered cold and dynamically passive. The specific scenario realized depends on the beam's directed velocity and the density ratio of beam to background electrons.

If the directed beam velocity exceeds the thermal velocity by at least a few times, then under typical conditions, even a relatively low beam density leads to a plasma wave growth rate that significantly exceeds that of the filamentation instability. Consequently, Langmuir turbulence of short-wavelength quasi-electrostatic fields develops much earlier than the Weibel turbulence of longer-wavelength quasi-magnetostatic currents. In this case, quasilinear calculations show that Langmuir turbulence, which flattens the electron velocity distribution between the beam and background components, can noticeably reduce the anisotropy parameter $A$ of the overall distribution. This, in turn, slows the exponential growth of filamentation modes, lowers their saturation level during further instability development, and alters the spectral evolution. Nevertheless, Langmuir turbulence cannot effectively suppress the development of filamentation turbulence. Moreover, the decay rate of Langmuir turbulence increases significantly, while the broadening of its spectrum is strongly suppressed due to the quasilinear influence of the developed magnetic turbulence.

If the beam velocity is only slightly above the thermal velocity—by a factor of two or less—then for relatively low beam densities, the growth rate of Langmuir modes becomes comparable to or even smaller than that of filamentation modes. As a result, Langmuir turbulence may not reach its saturation level and is rapidly suppressed by the growing magnetic turbulence. This suppression is primarily due to the quasilinear deformation and smoothing of the velocity distribution function in the resonant region with Langmuir waves. In other words, filamentation turbulence quasilinearly suppresses Langmuir turbulence.

The quasilinear analysis of the interaction between Langmuir and Weibel turbulence is important for identifying additional mechanisms of mutual impact between quasi-electrostatic and quasi-magnetostatic fluctuation spectra in beam-plasma systems. In addition to existing studies of three- and four-wave interactions among Langmuir~\cite{Kasaba2001,Ziebell2008} and Weibel~\cite{Garasev2021,Kuznetsov2025a} modes treated separately, the mutual interaction between filamentation and two-stream modes is of interest. In particular, such interaction may play a significant role in the process of anomalous scattering of electrons, resonant with Langmuir waves, by magnetic field fluctuations~\cite{Fleishman2013,Medvedev2017}. At high levels of turbulence saturation, when the turbulence can no longer be considered weak, a promising direction is to study the competition between quasilinear and nonlinear interactions of the two turbulence types. Specifically, the impact of quasi-electrostatic fields on current filaments (Z-pinches) and the effect of magnetic fluctuations on Langmuir solitons are relevant for understanding. Furthermore, anomalous collisions~--- caused by electron scattering on turbulence in a collisionless plasma~--- constitute a nonlinear mechanism for modifying the growth rates of Langmuir and filamentation modes, and therefore serve as another channel for mutual interaction between their spectra.

Finally, an important factor in the development and mutual impact of the mentioned turbulence spectra is the commonly present temperature difference between the background and the beam, as well as the possible presence of relativistic particles, both of which were neglected in the present study. For instance, a temperature difference between the background and beam allows for charge separation, which, if properly accounted for, reduces the growth rate of filamentation modes and the saturation level of magnetic turbulence~\cite{Tzoufras2006,Hao2008}. For a wide class of relativistic velocity distributions, the most unstable modes are oblique, with wave vectors making angles with the beam direction that differ significantly from both $0^\circ$ and $90^\circ$~\cite{Bret2004,Bret2010}. These and other factors can impact the quasilinear interaction between various types of turbulence and warrant further investigation. In astrophysical applications, the relevance of specific nonlinear effects that complement the developed quasilinear description of the joint evolution of Langmuir and Weibel turbulence is determined by the properties of nonequilibrium magnetoactive plasma in particular cosmic environments and lies beyond the scope of this study.

This research was supported by the Theoretical Physics and Mathematics Advancement Foundation “BASIS” (project no. 24-1-5-94-1). Numerical calculations were performed using the supercomputing facilities of the Keldysh Institute of Applied Mathematics of Russian Academy of Sciences. The source code used in this study is available at: .

\printbibliography
\end{document}